\documentclass[aps,prd,preprint]{revtex4-1}

\usepackage{graphicx,graphics,color}
\usepackage{float}

\usepackage{amsmath,amssymb}
\usepackage{mathrsfs}

\begin{document}

\title{Role of Quarks in Nuclear Structure}

\author{A. W. Thomas}
\affiliation{CoEPP and CSSM, Department of Physics, University of Adelaide,
	Adelaide SA 5005, Australia}



\date{\today}
\maketitle

\section{Introduction}
\label{sec:Intro}
Rutherford's discovery, more than a century ago, of the tiny nucleus at the heart of an atom~\cite{Rutherford:1911zz} created enormous problems for the physics of the day. With the electron and proton the only known building blocks and the Coulomb force the only known interaction, it was impossible to make such a compact object. The discovery of the neutron some 20 years later changed the game and suddenly one could envisage a new strong force between neutrons and protons that could bind them into nuclei. Since the 1930s a large part of nuclear theory has been concerned with deriving two- and three-nucleon forces and solving the many-body problem with those forces. Needless to say this approach has produced some impressive results.

The traditional approach which was just outlined treats the neutron and proton as elementary particles. Yet it was realized very early, from the anomalous magnetic moments of the proton and neutron, that these were not simple Dirac particles but had internal structure. Experiments with electron scattering in the 50s and 60s demonstrated that, far from being point-like, the proton and neutron are extended objects roughly 1 fermi (fm) in 
radius~\cite{Hofstadter:1957wk}. Not long after this Gell Mann and Zweig suggested that the dozens of strongly interacting ``elementary particles'' discovered by the early 60s might be composed of just three truly elementary particles called quarks (or aces)~\cite{GellMann:1981ph,Zweig:1981pd}. That this was the correct physical picture and not simply a mathematical artifice was made clear by the discovery of scaling in deep inelastic scattering and the later discovery of the $J/\Psi$. The rigorous basis for a theoretical description of the strong force as a local gauge theory, called Quantum Chromodynamics (QCD), built upon ``color'', with quarks bound by the exchange of gluons, was also developed in the early 70s~\cite{Fritzsch:1973pi}.

Like the famous electroweak theory, QCD is a local gauge theory. In Quantum Electrodynamics the underlying gauge principle is that one should be able to redefine the phase of a charged particle at each space-time point without changing the theory. This naturally leads to the existence of massless photons and all of the phenomena of electromagnetism. In the case of QCD the premise is that one should be able to arbitrarily change the color of a quark at each space-time point without changing the mathematical form of the theory. In this case the force carriers, or gluons,  also have spin-1 like the photon, but unlike the photon they carry charge and interact with each other through 3- and 4-gluon vertices. These interactions are a direct consequence of the fact that the generators of SU(3)-color do not commute and the consequences are profound. One can show rigorously that the theory is `asymptotically free'. That is, the forces between quarks decrease logarithmically at short-distance or, equivalently, high energy. 

In addition, QCD `confines' quarks, so that no-one has ever observed an isolated quark. Naively this may be understood in terms of a gluonic flux-tube of constant energy density which forms between quarks and persists no matter how far one takes them apart. The corresponding force is huge, of order 10 tonnes, and as the tube grows in length so does the energy tied up in this `string'. While this simple picture provides a natural explanation of confinement, because it would require infinite energy to separate an individual quark, the existence of light quarks,  which can break the flux-tube through the creation of a quark-anti-quark pair, adds a serious complication.

The remarkable strength of the force between quarks at the typical separations involved in hadrons means that so far the only rigorous way to calculate the properties of hadrons within QCD is to solve the equations numerically on a Euclidean space-time lattice. This has led to remarkable success for the masses of ground state hadrons~\cite{Durr:2008zz} and in the case of the nucleon for basic properties such as its electroweak form factors~\cite{Shintani:2018ozy,Stokes:2018emx}. For excited hadrons there remain challenges but there have already been some remarkable achievements. For example, we now know that the $\Lambda(1405)$, which caused so much trouble for the quark model is predominantly an anti-kaon-nucleon ($\bar{K} N$) bound state~\cite{Hall:2014uca}, rather than a traditional three-quark state. It also appears likely that the remarkably low mass positive parity excitation known as the Roper resonance is composite~\cite{Liu:2016uzk}, although this remains controversial for the moment~\cite{Burkert:2019bhp}.

Given this development of a far deeper understanding of the strong force in terms of quarks and gluons, with the protons and neutrons as relatively large composite objects, it may seem surprising that more attention has not been paid to the role of QCD and its more fundamental degrees of freedom in nuclear theory. Certainly there were brave attempts to use quark models to calculate the short-distance nucleon-nucleon (NN) force~\cite{Warke:1980jq,Detar:1978eu,Faessler:1982ik,Harvey:1980rva,Henley:1983ag,Oka:1981ri,Liberman:1977qs,Kalashnikova:1984qw} but tackling nuclear properties has rarely been attempted. In the early 80s, a remarkable discovery at CERN, the so-called EMC effect, named after the European Muon Collaboration, showed that the valence quark structure in a nucleus was severely modified compared with a free nucleon~\cite{Aubert:1983xm} but this has tended to lie outside the focus of nuclear theory. 

The plan of this overview is that we first outline the most commony used approaches to a QCD inspired theory of nuclear structure. This discussion starts with effective-field-theory, which uses the chiral symmetry of QCD to guide a systematic treatment at the level of nucleons and pions. Next we briefly outline the direct calculations of nuclear systems using lattice QCD. The latter constitute the only truly QCD based treatment of the problem. Both of these approaches will be treated in more detail elsewhere in this encyclopedia. Following those introductory remarks, we present a general discussion of the various views on whether or not an explicit treatment of the quark degrees of freedom is necessary in the quest to understand nuclear structure. We then describe an approach based upon the modification of nucleon structure in the presence of the strong scalar mean fields known to occur in a nuclear medium. This includes the derivation of a very effective energy density functional which has been successfully applied across the periodic table. Because this approach amounts to a new paradigm for nuclear theory, it is vital to find ways to test the predictions experimentally. We review a number of experiments that either have been performed or which may be expected to report results in the near future, including a precise measurement of the Coulomb sum rule for a variety of nuclei and, of course, the EMC effect and new variations on that. 

\section{Part I}
\label{sec:part1}
In this section we introduce several modern approaches to nuclear theory. The first of these is usually labelled effective-field-theory. Here the symmetries of QCD are encapsulated in an effective Lagrangian density in which the degrees of freedom are pions and nucleons. We explain why this has been considered such a promising approach and describe some of its successes. When it comes to tackling nuclear structure in terms of QCD itself, the only rigorous approach currently known is to use the techniques of lattice QCD. For the moment this approach is limited by technical issues, such as lattice size, rapidly increasing numbers of Wick contractions and so on. Nevertheless, some interesting results have already been reported at quark masses above the physical values, where the computing time is considerably reduced.

However, the main topic considered here is the approach in which one replaces point-like nucleons by extended objects with structure described by some quark model -- such as the MIT bag-model or the model of Nambu and Jona-Lasinio. One then takes into account the effect on that internal structure of exposing the confined quarks to the strong Lorentz-scalar mean fields known to appear in a nuclear medium. We shall see that the consequences of the predicted changes in nucleon structure in-medium are profound. For example, it provides a completely new mechanism to understand nuclear saturation. We explain the background to this approach and a number of examples of where it has been applied to nuclear phenomena.

\subsection{Effective-field-theory}
The clearest statement of why any explicit consideration of quark degrees of freedom might be ignored was made by Weinberg~\cite{Weinberg:1990rz}. One knows that QCD leads to extremely strong forces which confine quarks inside a hadron. Furthermore, one knows that chiral symmetry is spontaneously broken in QCD, with the very low mass of the pion identifying it as a pseudo-Goldstone boson (with its mass proportional to the square root of the quark mass) and the broken symmetry leading to the generation of relatively large consituent quark masses. Weinberg argued that the very low scale of nuclear binding, typically 8 MeV per nucleon, compared with the mass of the nucleon at 1 GeV, implied a ``separation of scales''. This suggested that one could build a non-relativistic, effective-field-theory (EFT) description of atomic nuclei, equivalent to QCD, based upon the symmetries of QCD, especially spontaneously broken chiral symmetry. Taking the EFT degrees of freedom as nucleons and pions, this program has indeed proven very successful~\cite{Entem:2017gor,Hu:2016nkw,Binder:2015mbz,Machleidt:2011zz,Furnstahl:2015rha,Hammer:2017tjm,Machleidt:2016rvv}, although in practice the number of parameters needed to describe NN scattering data is of order 30, comparable with phenomenological models~\cite{Wiringa:1994wb,Stoks:1993zz,Machleidt:1987hj}, which also accurately describe that data up to laboratory energies around 300 MeV. 

The EFT approach starts with a local Lagrangian density involving nucleons and pions and including all interactions consistent with chiral symmetry allowed to a given order (${\cal{O}}(p^2), \, {\cal{O}}(p^4)$, etc. ), in powers of momentum or pion mass, within some counting scheme~\cite{Kaplan:1998tg}. In general the interactions are not renormalizable, so that counter terms must be introduced order by order as each new divergence is regulated. The finite counter terms are then adjusted so that the theory gives the best possible fit to experimental data at the given order. The number of counter terms needed grows rapidly with the order to which one works. 

Having determined the NN force to a given order one can then solve the many-body problem to calculate the properties of atomic nuclei, with the main focus thus far on few-body systems~\cite{Binder:2015mbz} up to A=12. As with all other approaches which start with two-nucleon interactions, the disagreements with observed nuclear properties strongly suggested that something was missing. This led to the order by order calculation of a three-body force, again with counter terms chosen to fit key experimental data.

\subsection{Lattice QCD}
As discussed earlier, the only model independent way to use QCD to calculate hadron properties directly is lattice QCD. Given the success in calculating the ground state properties of hadrons it is natural that the lattice technique has been extended to nuclear systems. Several groups have calculated the phase shifts for baryon-baryon scattering using lattice QCD. This is numerically more challenging than solving the case of a single hadron because of the much larger spatial dimension of the lattice needed. Much of the work so far has involved larger quark masses than those found in nature, because of the  typical constraint that $m_\pi L$ (the pion mass times the size of the lattice ) should be considerably larger than one.

The HAL QCD collaboration has derived an effective local potential between baryons of strangeness $0, \, -1$ and $-2$, by calculating the corresponding correlation functions in lattice QCD~\cite{Nemura:2008sp,Murano:2011nz,Iritani:2018zbt}. Criticism of the method has focussed on the degree of model dependence of this approach but the results have been impressive. The aim of this approach is to derive the corresponding two-body potentials in this way and then use them in the nuclear many-body problem.

Rather than going through some $NN$ force, the NPLQCD collaboration has focussed its efforts on direct lattice QCD calculations of nuclear properties~\cite{Beane:2014ora,Orginos:2015aya,Winter:2017bfs,Chang:2017eiq,Chang:2015qxa,Beane:2010em}. So far this work has necessarily involved large quark masses, with the corresponding pion mass (recall $m_\pi^2 \propto m_q$ ) above 400 MeV. Nevertheless, even at such large masses the results for simple nuclear properties have been impressive. For example, the nuclear magnetic moments,  scaled by the corresponding nucleon mass, show a remarkable level of agreement with the experimental values for $d, \, ^3H$ and $^3He$~\cite{Beane:2014ora}. A major challenge for this approach, apart from the need for ever larger lattice volumes as the pion mass is lowered, is the rapid increase in the number of Wick contractions needed as the atomic number rises. This challenge has already led to the development of innovative methods of handling the problem~\cite{Detmold:2019fbk,Detmold:2012eu}.

Although the lattice approach to nuclear structure is still at a very early stage, there is no doubt that it will generate important new results over the next decade. The critical issue will be the extent to which these calculations will not just lead to better numerical results for nuclear properties but will actually allow us to develop new physical insight into the role of QCD in nuclear structure. As with hadron properties~\cite{Thomas:2002sj}, lattice studies of nuclear properties as a function of pion mass provide a powerful tool for seeking such insights.

\subsection{Medium modification of hadron properties}
Our main focus concerns an alternative approach which starts with the dispersion relation analysis of NN scattering by the Paris~\cite{Lacombe:1980gy} and Stony Brook~\cite{Brown:1978bx} groups in the mid-70s. That work demonstrated in a clear, model independent way that the dominant, intermediate range, attractive NN force~ was an iso-scalar Lorentz-scalar. In boson-exchange models of the NN force~\cite{Bryan:1964zzb,Green:1967yxm} this component has traditionally been described in terms of the exchange of a $\sigma$ meson, with the short-distance repulsion between nucleons described by the exchange of a 
Lorentz-vector meson, 
the $\omega$~\cite{Machleidt:1987hj}. The exchange of these mesons leads to large mean-fields in a nuclear medium~\cite{Horowitz:1982kp,Horowitz:1984sa}, albeit with an almost exact cancellation between them,  leading to the observed tiny binding energy per nucleon noted earlier. A number of researchers built upon these observations to develop a relativistic mean-field description of the nuclear many-body problem, such as Quantum Hadro-dynamics (QHD)~\cite{Serot:1979cc,Serot:1984ey,Horowitz:1982kp,Horowitz:1984sa}, in which the interactions of nucleons were described with a local Lagrangian density involving point-like nucleons interacting (at least in the simplest version) through the exchange of $\sigma$ and $\omega$ mesons.

Building on the success of QHD in a modern context, where the underlying theory of the strong force is QCD, one must ask whether the internal structure of the nucleon might reasonably be expected to remain unchanged in an environment with Lorentz-scalar and -vector mean-fields for which the strength can be a significant part of the mass of the nucleon itself. In our view the answer is clearly no. Whereas the coupling of a confined quark to a scalar field shifts its mass and thus changes the internal dynamics of the hadron, the coupling of the time component of a vector field (the only component which develops a mean-field expectation value in a medium at rest) to a quark simply shifts its energy, with no change in the underlying dynamics. For example, in the MIT 
bag-model~\cite{DeGrand:1975cf}, where this idea was first developed by Guichon~\cite{Guichon:1987jp} (see also Ref.~\cite{Fleck:1990td}), the scalar field shifts the effective quark mass entering the Dirac equation for the confined quark inside a bag to negative values as large as 100-200 MeV. This significantly modifies the 
bag-model wave-function, with the lower component being enhanced. 

The qualitatively different effects of the large Lorentz-scalar and -vector mean-field potentials on the internal structure of a hadron suggest immediately where the `separation of scale' arguments of the proponents of an effective-field-theory approach, based upon nucleon and pion degrees of freedom, might fail. Clearly the typical mean scalar field strength in an atomic nucleus is comparable to the energy required to excite a nucleon and one can no longer ignore such excitations.

The significance of the enhancement of the lower component of the Dirac wave-function of the confined valence quark should not be underestimated. For the nucleon as a whole, the effective nucleon-$\sigma$ coupling constant, $g_{\sigma N}$, is proportional to the integral of the upper Dirac component of the quark 
wave-function squared minus the square of the lower component. Thus the change in the internal structure of the nucleon naturally leads to a reduction of $g_{\sigma N}$ with increasing density. The Lorentz-vector character of the repulsive $\omega$-nucleon coupling means that it is density independent. {\em This provides an interesting new, natural mechanism for the saturation of nuclear matter, with the vector repulsion growing linearly with density while the growth of the scalar attraction is suppressed by the change in internal structure.} Indeed, this mechanism is sufficient to saturate nuclear matter even if we neglect the kinetic energy of the nucleons, unlike QHD where the relativistic correction to the nucleon kinetic energy provides the saturation mechanism.

Mathematically, the change of the $\sigma N$ coupling with increasing scalar-field strength requires a self-consistent solution of the field equations. The resulting effective mass of the nucleon may be written as
\begin{equation}
M_N^* = M_N - g_{\sigma N}(\sigma) \sigma 
\label{eq:Mnstar}
\end{equation}
where the nucleon-$\sigma$ coupling may be expressed as
\begin{equation}
g_{\sigma N}(\sigma) = g_{\sigma N}(0) [1 \, - \, \frac{d}{2} (g_{\sigma N}(0) \sigma)] \, .
\label{eq:d}
\end{equation}
By analogy with electromagnetism, where the electric polarizability characterizes the tendency of the electrons in an atom to rearrange to oppose an applied electric field, $d$ is called the {\em scalar polarizability}. In the bag-model, $d$ is approximately $0.2 \, R$, with $R$ the bag radius.

While the development of this approach, known as the `quark-meson coupling (QMC)' 
model~\cite{Guichon:1995ue,Saito:1994ki,Guichon:2018uew,Saito:1995up,Saito:1996yb}, was based upon the MIT bag-model for hadron structure, it should be clear that the general features are not model dependent. Indeed, a formulation based upon the model of Nambu and Jona-Lasinio (NJL)~\cite{Nambu:1961tp}, which is covariant and respects chiral symmetry, leads to the same mechanism for the saturation of nuclear matter~\cite{Bentz:2001vc}. 

As we discuss later, this approach to nuclear structure has been used to derive an energy density functional 
(EDF)~\cite{Guichon:2006er}, with a form a little more complicated than the usual Skyrme forces~\cite{Krewald:1977oev} but which has nevertheless been applied to nuclear structure with some success. In particular, the quality of the overall description of nuclear binding and sizes across the periodic table was found to be comparable with modern Skyrme forces~\cite{Kortelainen:2010hv} but with considerably fewer parameters.

{\em Without meaning to diminish the importance of the results for gross nuclear properties, we stress the change of paradigm that it represents.} Rather than neutrons and protons occupying shell model orbits in a nucleus, those orbits are occupied by clusters of quarks with the quantum numbers of nucleons but whose internal structure has adjusted to the local scalar mean-field.
This represents a fundamental shift in our conceptual understanding of the structure of atomic nuclei. It is therefore crucial to find ways to test the new paradigm empirically and much of the next section will be devoted to methods that have been proposed to do that.

\section{Part II}
\label{sec:part2}
In this section we focus on the QMC model for which the motivation was set out in Sec.IIC. We stress that this is indeed a model, not QCD itself. Nevertheless, as we have explained, it does allow us to examine the consequences of the fact that the structure of composite nucleons may be expected to change in a nucleus with its large Lorentz-scalar and -vector potentials.
Starting at the quark level it has already been possible to achieve a number of very interesting theoretical results. Here we begin with the derivation of an EDF which has been applied to the calculation of the properties of atomic nuclei across the periodic table. Key results are described, with a particular emphasis on the success for superheavy nuclei. Because the model is derived at the quark level, calculations of the properties of other hadrons, such as, for example,  the $\eta$ meson, hyperons and heavy quark systems, when imbedded in nuclear matter involve {\em no new parameters}. 

\subsection{Energy density functional}
 Since the pioneering work of Vautherin and Brink~\cite{Vautherin:1971aw} in the 1970s, much of the study of nuclear properties has been based upon Hartree-Fock calculations starting from an EDF~\cite{Bender:2003jk,Kortelainen:2010hv,Kortelainen:2013faa,Klupfel:2008af,Bogner:2013pxa}. The so-called Skyrme forces have been much explored in this context, with hundreds of such forces each having ten or more parameters tuned to a particular set of nuclear properties. Although the QMC model is built upon the fundamental difference between a Lorentz-scalar and -vector potential and this is vital if one wishes to look for changes in the properties of bound nucleons, in order to calculate nuclear properties it is far more efficient to derive an equivalent, non-relativistic EDF and work with that.

In the first attempt to connect the QMC model to typical nuclear structure calculations, Guichon and 
Thomas~\cite{Guichon:2004xg} derived a local, density independent force of the Skyrme type as an approximation to the underlying model. The scalar polarizability, which as we explained earlier characterizes the response of the internal structure of the nucleon to the applied scalar field, was shown to lead naturally to a three-body force, with a strength predicted by the model. A comparison with phenomenological Skyrme forces of this type showed that the predicted three-body force had a strength remarkably close to the phenomenological value.

While this first step was promising, the derived Skyrme force was at best an approximate representation of the underlying model. Furthermore, modern studies of nuclear structure have abandonned the use of local, density independent forces with two- and three-body terms in favour of density dependent forces. Of course, the density dependence is equivalent physically to introducing a three-body force but the density dependent forces have proven more practical. This led Guichon {\em et al.}~\cite{Guichon:2006er} to derive an EDF with a dependence on density beyond quadratic that is equivalent to the underlying quark-based theory. The EDF had the form
\begin{equation}
<H(\vec{r})>=\rho M+\frac{\tau}{2M}+{\cal H}_{0}+{\cal H}_{3}+
{\cal H}_{eff}+{\cal H}_{fin}+{\cal H}_{SO}
\label{eq:35}
\end{equation}
where
\begin{eqnarray}
{\cal H}_{0}+{\cal H}_{3} & = & {\rho}^{2}\,\left[\frac{-3\,{G_{\rho}}}{32}+
\frac{{G_{\sigma}}}{8\,{\left(1+d\,\rho\,{G_{\sigma}}\right)}^{3}}-
\frac{{G_{\sigma}}}{2\,\left(1+d\,\rho\,{G_{\sigma}}\right)}+
\frac{3\,{G_{\omega}}}{8}\right]+\\
 &  & {\left({{\rho}_{n}}-{{\rho}_{p}}\right)}^{2}
\left[\frac{5\,{G_{\rho}}}{32}+\frac{{G_{\sigma}}}{8\,{\left(1+d\,\rho\,{G_{\sigma}}\right)}^{3}}-\frac{{G_{\omega}}}{8}\right],
\label{eq:40}\\
{\cal H}_{eff} & = & \left[\left(\frac{{G_{\rho}}}{8\,{m_{\rho}}^{2}}-\frac{{G_{\sigma}}}{2\,{m_{\sigma}}^{2}}+\frac{{G_{\omega}}}{2\,{m_{\omega}}^{2}}+
\frac{{G_{\sigma}}}{4\,{M_{N}}^{2}}\right)\,{{\rho}_{n}}+
\left(\frac{{G_{\rho}}}{4\,{m_{\rho}}^{2}}+
\frac{{G_{\sigma}}}{2\,{M_{N}}^{2}}\right)\,{{\rho}_{p}}\right]\,{{\tau}_{n}}
\nonumber \\
&  & +p\leftrightarrow n,
\label{eq:41}\\
{\cal H}_{fin} & = & \left[\left(\frac{3\,{G_{\rho}}}{32\,{m_{\rho}}^{2}}-\frac{3\,{G_{\sigma}}}{8\,{m_{\sigma}}^{2}}+\frac{3\,{G_{\omega}}}{8\,{m_{\omega}}^{2}}-\frac{{G_{\sigma}}}{8\,{M_{N}}^{2}}\right)\,{{\rho}_{n}}\right.\\
 &  & +\left.\left(\frac{-3\,{G_{\rho}}}{16\,{m_{\rho}}^{2}}-
\frac{{G_{\sigma}}}{2\,{m_{\sigma}}^{2}}+
\frac{{G_{\omega}}}{2\,{m_{\omega}}^{2}}-
\frac{{G_{\sigma}}}{4\,{M_{N}}^{2}}\right)\,{{\rho}_{p}}\right]{{\nabla}^{2}}({{\rho}_{n}})+p\leftrightarrow n,
\label{eq:42}\\
{\cal H}_{SO} & = & \nabla\cdot{J_{n}}\left[\left(\frac{-3\,{G_{\sigma}}}{8\,{M_{N}}^{2}}-\frac{3\,{G_{\omega}}\,\left(-1+2\,{{\mu}_{s}}\right)}{8\,{M_{N}}^{2}}-\frac{3\,{G_{\rho}}\,\left(-1+2\,{{\mu}_{v}}\right)}{32\,{M_{N}}^{2}}\right)\,{{\rho}_{n}}\right.\\
 &  & +\left.\left(\frac{-{G_{\sigma}}}{4\,{M_{N}}^{2}}+\frac{{G_{\omega}}\,\left(1-2\,{{\mu}_{s}}\right)}{4\,{M_{N}}^{2}}\right)\,{{\rho}_{p}}\right]+p\leftrightarrow n \, .
\label{eq:edf}
\end{eqnarray}
Here $G_i = g_i^2/m_i^2$, where $m_i$ is the meson mass for $i=\sigma, \omega, \rho$ and $g_i$ is the free space meson-nucleon coupling constant. The proton and neutron densities are respectively, $\rho_p$ and $\rho_n$, while $\tau_p$ and $\tau_n$ are the usual kinetic energy operators and $\mu_s$ and $\mu_v$ the isoscalar and isovector magnetic moments of the nucleon. These are calculated within the particular hadronic model in terms of the quark-meson couplings, which are the underlying fundamental parameters of the model. For simplicity and motivated by Zweig's rule, these mesons are taken to have no coupling to the strange quarks. 
\begin{center}
	\begin{figure}[H]
		\centering
		\includegraphics[angle=0,width=0.7\textwidth]{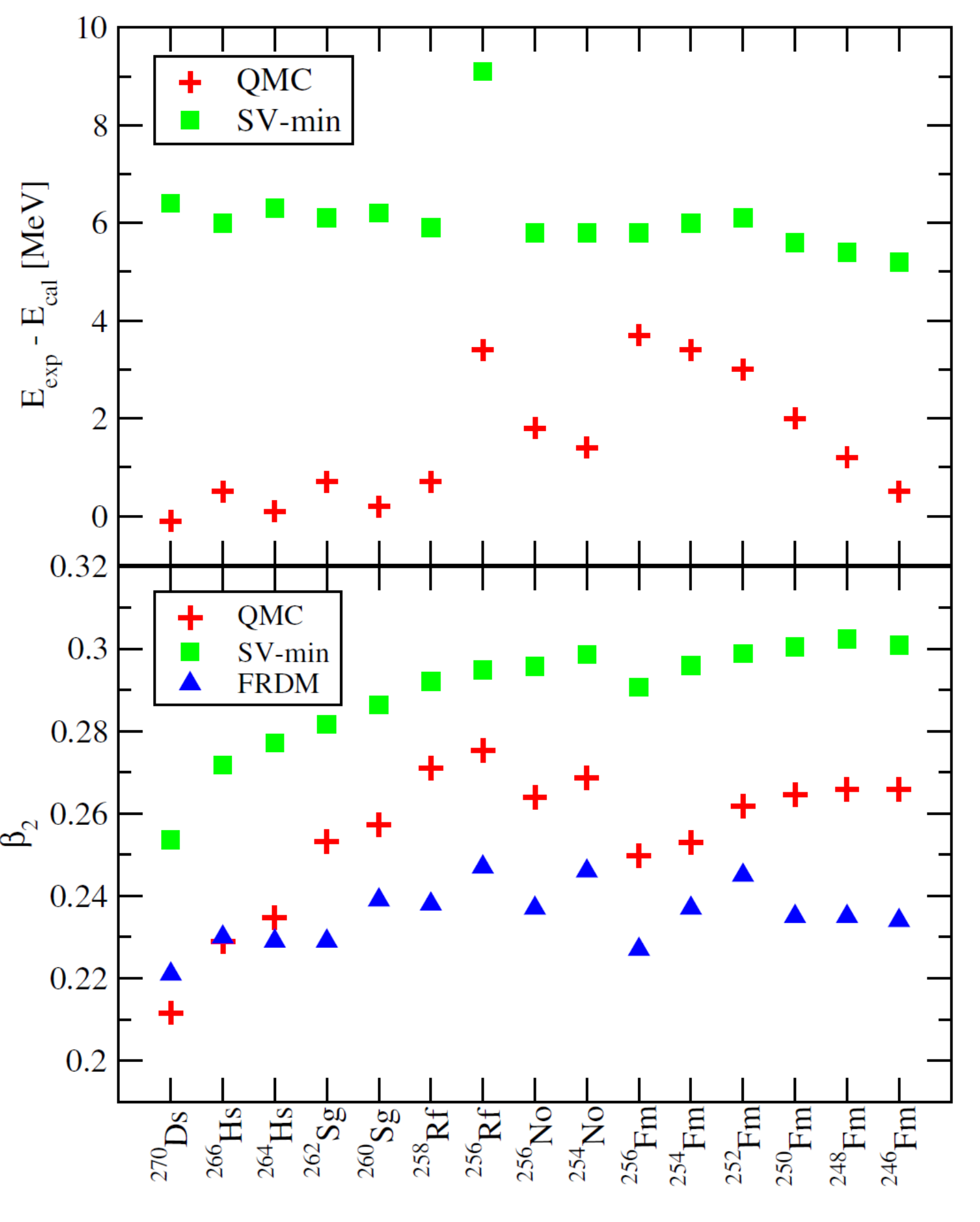}
		\caption{Comparison of the difference between the calculated and experimental total ground-state binding energies of superheavy nuclei calculated within the QMC model, as well as using the Skyrme force SV-min~\cite{Klupfel:2008af} and the finite-range droplet model (FRDM)~\cite{Moller:1993ed}. The deformations predicted for these nuclei are also illustrated -- from Ref.~\cite{Stone:2016qmi}.}
		\label{fig.BE}
	\end{figure}
\end{center}
%

The novel higher order density dependence appearing in Eq.~(\ref{eq:edf}) is a consequence of the self-consistent adjustment of the nucleon internal structure in-medium. Note that this behaviour would disappear in the limit where the scalar polarizability, $d$,  goes to zero. It is equivalent to the appearance of a repulsive three-body force.  We stress that the strength of this three-body force is predicted by the theory and does not involve new parameters. Similar forces appear if we introduce any other baryon into the medium, so that in dense nuclear matter, as one may find in a heavy neutron star where hyperons may be expected to appear~\cite{Glendenning:1982nc}, one naturally finds repulsive $HNN, HHN$ and $HHH$ forces, again with no new parameters~\cite{RikovskaStone:2006ta,Stone:2010jt}.
\begin{center}
	\begin{figure}[H]
		\centering
		\includegraphics[angle=0,width=0.7\textwidth]{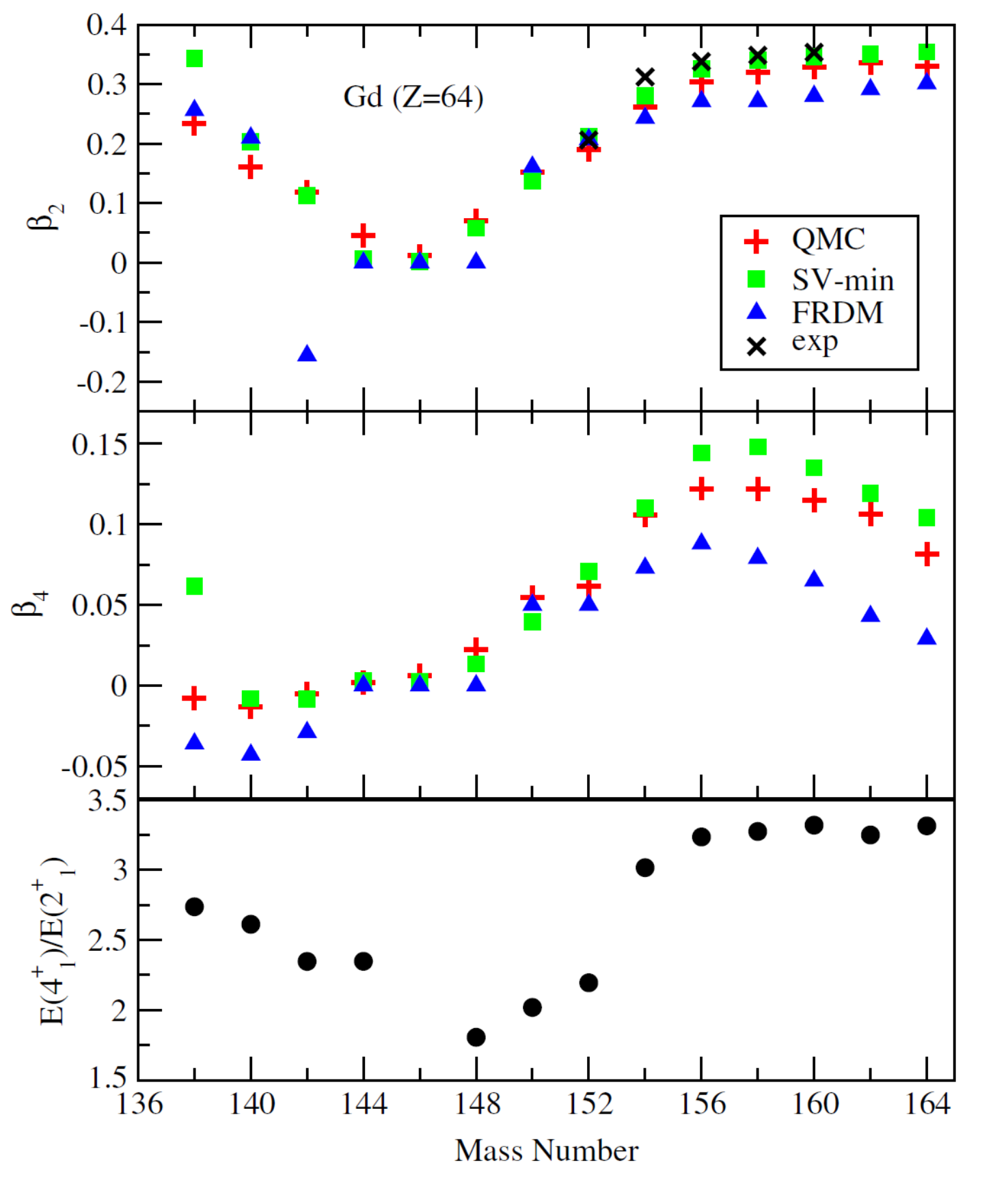}
		\caption{Deformation parameters of the $^{138-164}Gd$ isotopes -- from Ref.~\cite{Stone:2016qmi}. We observe the close agreement between the QMC model and the experimental deformation parameters where they are known. Where they are not known the predictions agree well with the results of the very successful FRDM model.}
		\label{fig.deformation}
	\end{figure}
\end{center}

The first systematic study of nuclear structure within this framework was carried out by 
Stone {\em et al.}~\cite{Stone:2016qmi} using the Hartree-Fock plus BCS code Skyax~\cite{Klupfel:2008af}. Their work omitted  the Fock terms arising from pion exchange, which tend to lower the nuclear incompressibility. Nevertheless, the results were very promising. The four parameters (apart from two pairing parameters) of the model, $G_{\sigma, \omega, \rho}$ and $m_\sigma$, were first constrained to reproduce the saturation energy and density of nuclear matter as well as the symmetry energy, within reasonable errors. Within that constraint they were then fit to the properties (primarily binding energy, charge radius, surface thickness and pairing gaps) of a large set of nuclei across the periodic table. (Note that $m_\sigma$ is treated as a parameter because it is not well determined empirically and suffers an ambiguity in the quantization procedure~\cite{Guichon:1995ue}.) The binding energies were reproduced with an rms error of just 0.36\%, compared with 0.24\% for SV-min~\cite{Klupfel:2008af} (which was fit to the same data set). For charge radii the rms errors were 
0.7\% and 0.5\%, respectively. Although the QMC model does not reproduce the data quite as well as SV-min, it is remarkable that SV-min involved more than twice as many parameters in the nuclear force.
\begin{center}
	\begin{figure}[H]
		\centering
		\includegraphics[angle=0,width=0.8\textwidth]{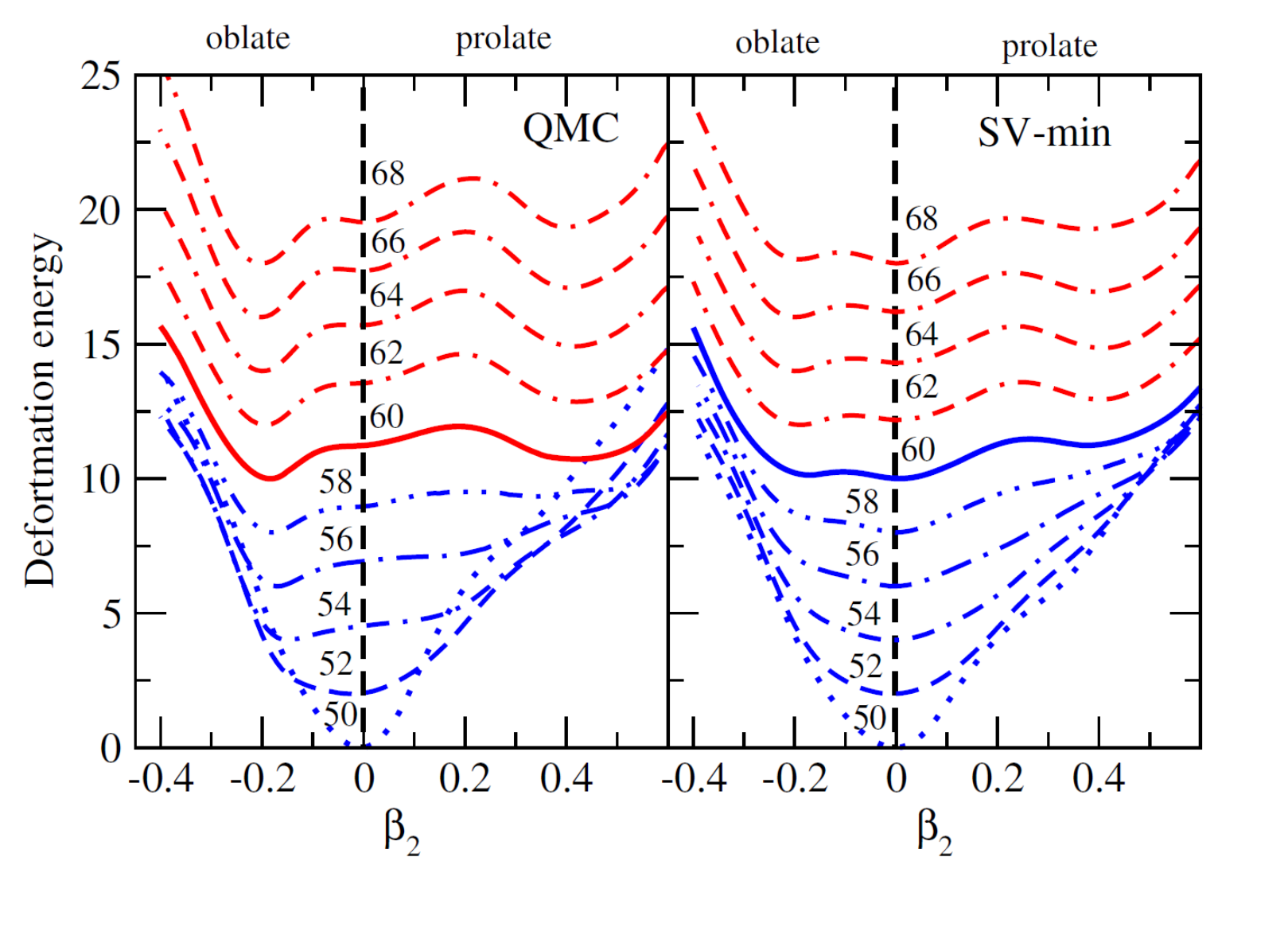}
		\caption{Illustration of the phenomenon of shape evolution for the $^{90-106}Zr$ isotopes as a function of neutron number obtained in constrained HF+BCS with QMC (left panel) and SV-min (right panel) EDF -- from Ref.~\cite{Stone:2016qmi}.  Deformation energies are displayed as a function of the quadrupole deformation parameter $\beta_2$. The same arbitrary constant has been added for each isotope in both panels for display purposes. The vertical dashed line indicates the spherical shape.
}
		\label{fig.coexistence}
	\end{figure}
\end{center}

Figure~\ref{fig.BE} illustrates the relatively small deviations in total binding energy (~1-2 MeV in typically 1500 MeV) between the predictions of the model for super-heavy nuclei, with $Z \in (100,110)$, {\em which were not part of the fit}. The rms deviation for this set is just 0.1\% in the QMC model, compared with 0.3\% for SV-min. This certainly provides encouragement to further explore the predictions of the model for super-heavy nuclei. In Fig.~\ref{fig.deformation} we show the deformation parameters for isotopes of Gd, as an illustration of the capacity of the model to deal with such phenomena. Where experimental data are shown the QMC model is in quite good agreement as are the predictions of FRDM. In those cases where there are no data we observe that the QMC calculations are relatively close to those of FRDM. As the latter is a very successful phenomenological approach, with many more parameters tuned to a wide range of experimental data, this level of agreement is an indication that the QMC model does reproduce the behaviour of the nuclear deformation quite well.  As a last illustration of the results of this first study, 
in Fig.~\ref{fig.coexistence} we show that the model reproduces the fascinating shape co-existence, that is the near degeneracy of the minima for spherical, oblate and prolate shapes, in nuclei in the region 
around $A=100$~\cite{Sotty:2015hya}.

\subsection{Binding of other hadrons}
Because the coupling constants entering the QMC model involve mesons coupling to the quarks, one can calculate the meson couplings to {\em any hadron} with {\em no additional} parameters. Thus, at mean-field level one can make unambiguous predictions for the binding of any hadron when immersed in nuclear matter. We refer to the work of Saito {\em et al.}~\cite{Saito:2005rv} for a comprehensive review.

\subsubsection{Mesons}
In mesons built of $u$ or $d$ quarks the repulsion associated with the $\omega$ mean field cancels, so that the meson necessarily feels a considerable attractive force. Thus within the QMC model one is naively led to expect the $\omega$ and $\rho$ mesons to be deeply bound in a nucleus. Unfortunately nature is a little more complicated than this. Both of these mesons experience strong absorption in a nuclear medium leading to two-nucleon emission. It is well known that the dispersive effect of such absorption is a repulsive interaction and so, in practice, it is difficult to extract unambiguous information from studies of these mesons in nuclei.

The $\eta$ and $\eta^\prime$ mesons are more promising~\cite{Bass:2018xmz} and there has been extensive work at COSY as well as ELSA, JLab, GrAAL, GSI and Mainz to look for signs of a sufficiently strong attraction that the meson may be able to bind to the nucleus~\cite{Krusche:2014ava,Metag:2017yuh,Witthauer:2017get,Kashevarov:2017kqb,Krusche:2012zza,Senderovich:2015lek,Friedrich:2016cms}. For the $\eta^\prime$, in particular, data on photoproduction at ELSA have suggested that even at nuclear matter density the width of a bound state should be around 20 MeV. This is sufficiently small to give one hope of being able to detect relatively narrow $\eta^\prime$-nucleus bound states.

Indeed, by studying the excitation function in $\eta^\prime$ photoproduction on a carbon target, the CBELSA/TAPS Collaboration reported that the real potential felt by an $\eta^\prime$ meson at nuclear matter density is attractive, with a depth of order $37 \pm 10 \pm10$ MeV~\cite{Nanova:2013fxl}. A later experiment on Nb found a similar attraction, namely $41 \pm10 \pm 15$ MeV~\cite{Nanova:2016cyn}. One would like to see the errors reduced with further study but these results are already very promising. The theoretical analysis of the $\eta-\eta^\prime$ system is complicated by the role of the axial anomaly and the mixing with the famous Witten-Veneziano flavour-singlet  gluonic 
state~\cite{Veneziano:1979ec}. Nevertheless, taking this into account within the context of the QMC model, Bass and 
Thomas~\cite{Bass:2005hn,Bass:2013nya} reported an attractive potential of this size well before the experiment.

\subsubsection{Hyperons}
When baryons are immersed in a nuclear medium one has to deal with both the scalar and vector mean field potentials and the cancellation between them means that the binding should be comparable with that of the nucleon. As we have explained, knowing the quark couplings to the $\sigma, \omega$ and $\rho$ mesons to the up and down quarks one can calculate the meson-baryon coupling constants without new parameters. Because of the extensive experimental studies 
of $\Lambda$ hypernuclei, this is a particularly interesting case. 

Studies of hypernuclei within the QMC model~\cite{Guichon:2008zz,Tsushima:1997cu} found the following key results.
\begin{itemize}
\item Firstly, by working at the quark level with the mesons coupled only to the light quarks the observed fact that the spin-orbit force for $\Lambda$ hyperons is very small was naturally explained. In particular, as the Lorentz-scalar and -vector potentials act only on the non-strange quarks, which have total spin zero in the $\Lambda$, there is no direct spin-orbit term. This leaves only a small term arising from Thomas precession of the whole system in which the scalar and vector contributions almost cancel.
\item Secondly, the binding of the $\Lambda$ hyperon in the $1s$-state of lead was within an MeV~\cite{Guichon:2008zz} of the experimental value, which is around 27 MeV~\cite{Ajimura:1994jb}.
\item In the early work the $\Sigma$ hyperon experienced significant binding, which contradicts the experimental evidence that the $\Sigma$ experiences repulsion in nuclear matter.
\end{itemize}

The solution to the latter problem is fascinating and illustrates the unexpected importance of working at the quark level. We recall first that the so-called hyperfine interaction plays an extremely important role in hadron spectroscopy. It is responsible for most of the 300 MeV splitting between the masses of the $N$ and $\Delta$ as well as the roughly 70 MeV splitting between the $\Sigma$ and $\Lambda$ hyperons~\cite{DeGrand:1975cf,Thomas:1982kv}. Gluon exchange between quarks induces a spin-spin interaction which is attractive for a spin-0 pair of quarks and repulsive for a spin-1 pair. It is inversely proportional to the product of the masses in a simple constituent quark model, or the eigenenergies in the bag-model. The fact that the light quarks have spin-0 in the $\Lambda$ and spin-1 in the $\Sigma$ then naturally explains why the latter is heavier. 

In a nuclear medium the effect of the mean scalar field is to lower the quark eigenenergies, which naturally increases the $\Sigma-\Lambda$ mass difference. The nett result in the more sophisticated implementation of the QMC model, which takes this into account~\cite{Guichon:2008zz}, is that while the $\Lambda$ hyperon is still bound by essentially the observed amount in Pb, the $\Sigma$ hyperon experiences extra repulsion in  nuclear matter, consistent with the phenomenological findings. In addition, in very dense matter, such as one might expect in the core of a heavy neutron star, the consequence of this enhancement of the hyperfine interaction in medium is that $\Sigma$ hyperons do not appear at all~\cite{RikovskaStone:2006ta} and nor do $\Delta$ baryons~\cite{Motta:2019ywl}. Instead, one expects to find $\Xi^-$ and $\Xi^0$ hyperons as well as $\Lambda$ hyperons. In this context it is encouraging that the first results from J-PARC concerning $\Xi$ 
hypernuclei~\cite{Kinbara:2019kyx} tend to provide some support for the small binding expected in the QMC model.

\section{Experimental tests of changing hadron structure}
Given the fundamental change in the picture of the atomic nucleus implied by starting at the quark level, it is vital to find experimental ways to test the concept. A number of possibilities have already been explored as we shall briefly explain. The first concerns the potential modification of the electric form factor of the proton, $G_E(Q^2)$. Here the challenge is that one cannot directly measure the form factor of the bound nucleon, rather one must measure a process such as quasi-elastic scattering which depends on it along with other details of the reaction dynamics. The two most promising experiments are the careful determination of the Coulomb sum rule for inelastic electron scattering from nuclei and the recoil polarization measurement which is sensitive to the ratio of the electric to the magnetic form factor of the proton. We discuss these examples below. We then turn to the predicted change of the axial form factor of the nucleon and its effect on neutrino scattering and double $\beta$-decay. Finally, we review the modification of nuclear structure functions by the European Muon Collaboration at CERN, known as the EMC effect.

\subsection{Coulomb sum rule}
Quasi-elastic scattering of an electron from a nucleus involves two response functions which depend on the three-momentum and the energy transfered~\cite{Wehrberger:1993zu}. The longitudinal response is particularly interesting in that its dependence on nucleon (as opposed to nuclear) properties is primarily on the electric form factor of the proton. If, as predicted in the QMC approach, the proton electric form factor were to be softer~\cite{Saito:1999bv}, this should be observable. Remarkably, experiments at the ALS at Saclay more than 30 years ago did suggest precisely 
this~\cite{Meziani:1984is}. 
%
	\begin{figure}[H]
		\centering
		\includegraphics[angle=0,width=0.9\textwidth]{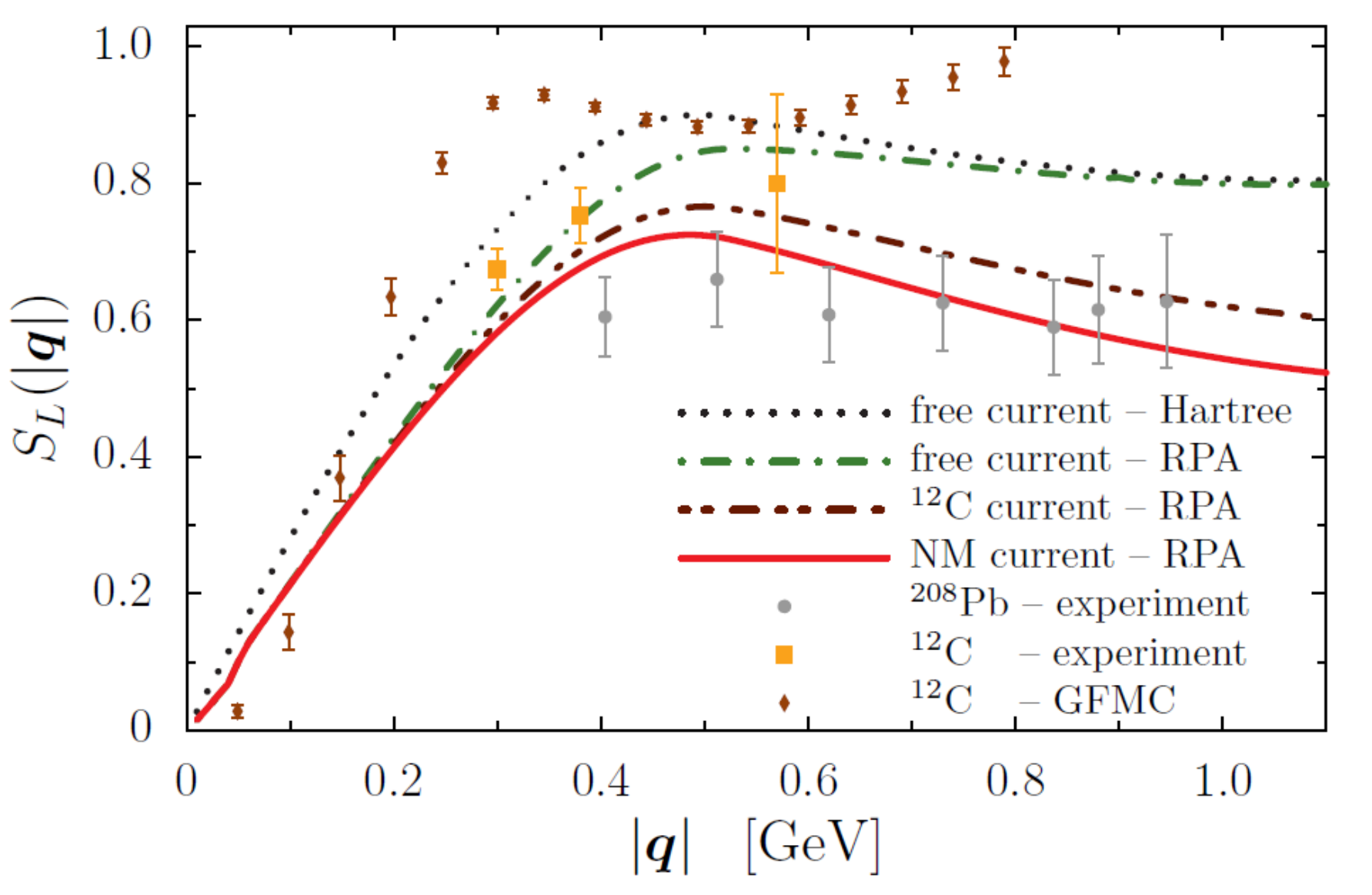}
		\caption{Coulomb sum rule determined at $\rho = 0, 0.1$ and 0.16 $fm^{-3}$, corresponding to a free nucleon current, a density typical of $^{12}$C; and nuclear matter saturation density --  from 
Ref.~\cite{Cloet:2015tha}. The data for $^{208}$Pb is from 
Refs.~\cite{Zghiche:1993xg,Morgenstern:2001jt} and for $^{12}$C from Ref.~\cite{Barreau:1983ht}, both without the relativistic correction factor of de Forest~\cite{DeForest:1984qe}. The Green's function Monte Carlo results are taken from Ref.~\cite{Lovato:2013cua}. The effects of relativity and the predicted modification of the proton electric form factor in-medium are both dramatic, with each tending to lower the Coulomb sum rule at large three-momentum transfer, $|\bf{q}|$, by as much as 20\%. RPA indicates that those calculations include the correlations arising within the random phase approximation.}
\label{fig.CSR}
	\end{figure}
%

Even earlier, McVoy and van Hove had proposed using the longitudinal response in a different way~\cite{McVoy:1962zz}. They observed that the quantity known as the Coulomb sum rule, obtained by integrating the longitudinal response over the energy transfer, after removing the free proton electric form factor and the charge of the nucleus, would tend to unity at large three-momentum. The physical idea is simple, one is simply counting the number of charged particles that can be knocked out if the three-momentum is much higher than the momentum of any bound proton.  Their proposal was that the rate at which the Coulomb sum approached  unity at large three-momentum transfer would provide information on short-range correlations.

Spurred by the importance of the potential change in the structure of a bound nucleon and the controversy over the first measurement, a JLab experiment was performed to very precisely map the longitudinal response over a wide enough range of energy to unambiguously determine the Coulomb sum rule and compare with the predictions of various nuclear models. Figure~\ref{fig.CSR} shows the data from the 1980s, with the C data not very conclusive but the Pb data suggesting a very significant suppression of the Coulomb sum rule below unity at large momentum transfer. The non-relativistic Green's function monte carlo calculation for C tends to unity at 0.8 GeV/c, while the relativistic calculations, based on the self-consistent change in proton structure predicted in-medium (at a nuclear matter density matching the average nuclear density), tend to reproduce the reduction in Pb quite well~\cite{Cloet:2015tha}.  More precise results from a recent JLAB experiment~\cite{JLab-E05} should soon be available. 

\subsection{Recoil polarization}
Another remarkable experiment aimed at testing the change in the proton electric form factor made full use of the capabilities of the JLab facilities~\cite{Strauch:2002wu}. If a proton knocked out of the nucleus can be detected in coincidence with a scattered electron {\em and} its polarization along and perpendicular  to the outgoing direction measured, $G_E/G_M$, the ratio of the electric and magnetic form factors, can be obtained. This technique was used to reveal the completely unexpected decrease of this ratio at high momentum transfer for the free proton~\cite{Jones:1999rz}, when earlier SLAC data using Rosenbluth separation~\cite{Thomas:2001kw} had suggested the ratio was near one, independent of the four-momentum transfer, $Q^2$. Because the measurement yields a ratio of form factors it was hoped that nuclear corrections might cancel to some extent.
\begin{center}
	\begin{figure}[H]
		\centering
		\includegraphics[angle=0,width=0.9\textwidth]{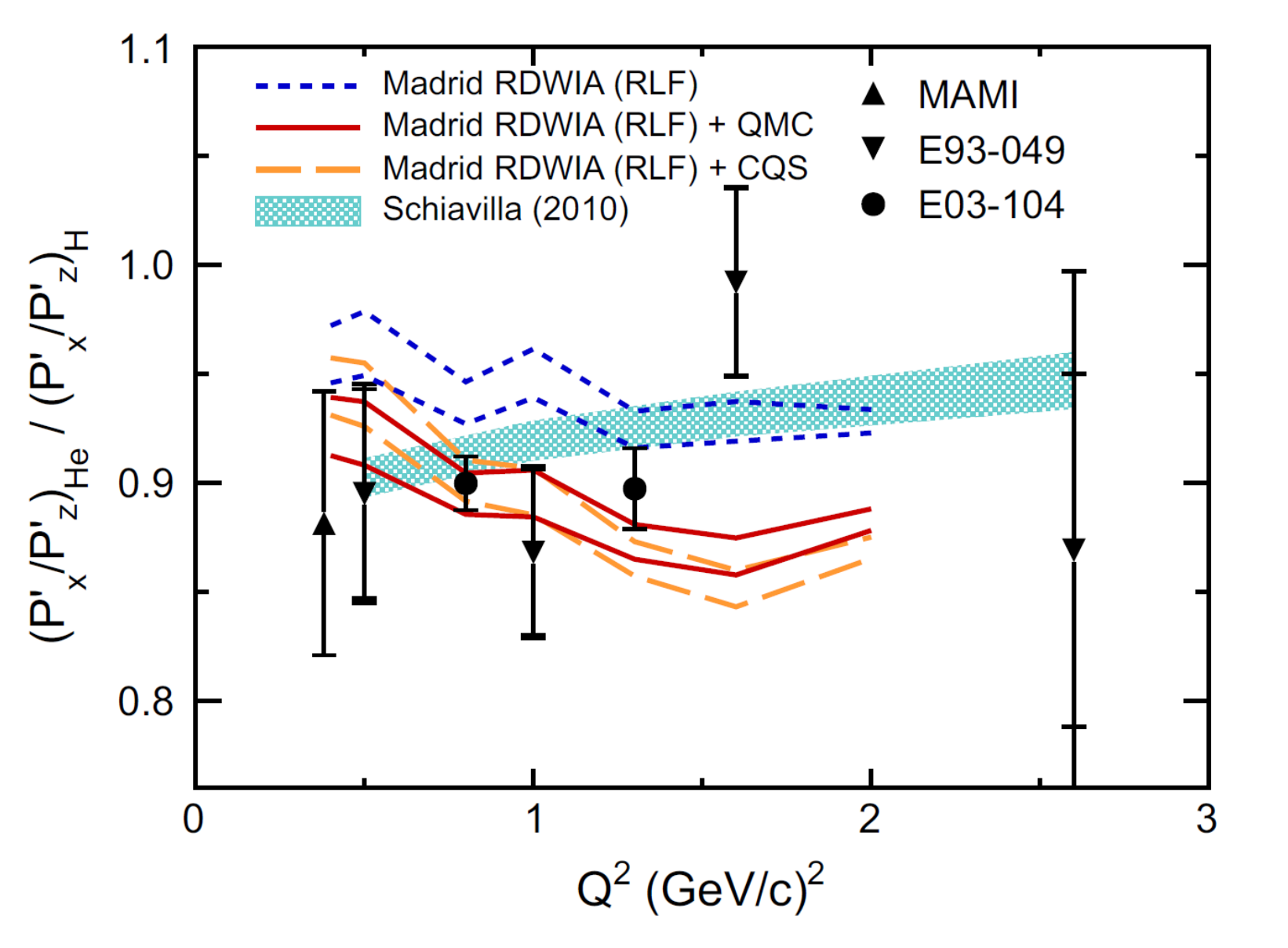}
		\caption{Polarization-transfer double ratio R as a function of $Q^2$ -- from Ref.~\cite{Strauch:2002wu}. 
The data are from Mainz~\cite{Dieterich:2000mu} and Jefferson Lab experiments   
E93-049~\cite{Strauch:2002wu} and 
E03-104~\cite{Paolone:2010qc}. The shaded area shows the results of calculations by 
Schiavilla {\em et al.}~\cite{Schiavilla:2004xa}, while the other curves are from the Madrid 
group~\cite{Udias:2000ig}. 
In-medium form factors calculated in the QMC model well before the 
experiment~\cite{Lu:1997mu} (solid curve, labelled QMC) were used in two of the Madrid calculations. }
\label{fig.strauch}
	\end{figure}
\end{center}

Figure~\ref{fig.strauch} shows the results of the JLab experiment for the ratio of the $G_E/G_M$ ratios in He and hydrogen in comparison with a number of theoretical calculations. The case of particular interest here is given by the red lines, labelled Madrid QMC. This lies significantly below the blue dashed band which has the nuclear distortion effects but no in-medium change in proton structure. The level of agreement is remarkable given that the medium modification corrections were calculated long before the experiment and were therefore a genuine prediction. The solid blue band, labelled Schiavilla~\cite{Schiavilla:2004xa}, was calculated after the data showed a discrepancy with calculations based on free-space form factors and includes a rather strong charge-exchange final-state interaction which has not been verified in independent experiments. 

\subsection{Axial form factor}
The axial form factor of the nucleon plays a major role in weak interactions. Any modification in-medium may be expected to be important in calculating neutrino-nucleus interactions, which are especially relevant to the interpretation of neutrino oscillation experiments~\cite{Hutauruk:2018cgu}. In double-beta decay, the lifetime goes roughly as the fourth power of the axial charge, which makes it a crucial parameter.

From the very beginning of the development of the quark-based approach to nuclear structure it was clear that the effect of a strong mean scalar field would be to reduce the axial charge, $g_A$. Recall that the reduction with density of the scalar coupling to the nucleon comes from the increase of the lower Dirac component of the quark wave-function, with that coupling proportional to the integral over $u^2 - l^2$, where $u$ and $l$ are the upper and lower components of the bound Dirac spinor. As $g_A$ goes like $u^2 - l^2/3$, there is a reduction in this charge but it is somewhat smaller than that for the scalar coupling. Indeed, at nuclear matter density one expects a reduction of order 10\%. In a recent study the changes in bound form factors were estimated to increase the neutrino mean free path by typically 15-20\% in nuclei and by as much as 40\% in the dense matter expected in the core of a neutron star~~\cite{Hutauruk:2018cgu}. These are important changes which need further careful investigation.

\subsubsection{$\Lambda$ beta-decay in-medium}
The decay mode $\Lambda \rightarrow p + e^- + \bar{\nu}_{e}$ is very rare but offers a clean probe of any potential change in weak form factors if the $\Lambda$ is immersed in a nuclear medium. 
Guichon and Thomas~\cite{Guichon:2017gbe} recently examined the change in the relevant $|\Delta S | = 1$ vector and axial vector form factors in a study encouraged by the possibility of testing predictions for a change in this partial decay width at J-PARC. The study carefully preserved the Ademollo-Gatto 
theorem~\cite{Ademollo:1964sr,foot} in the case of the vector form factor, finding a reduction of order 3-4\% at nuclear matter density, while for the axial form factor the reduction at that density was of order 8\%. Changes of this order of magnitude should be clearly visible as a change of the partial width for $\Lambda$ beta-decay in-medium.

\subsection{EMC effect}
It is now more than 30 years since the European Muon Collaboration (EMC) at CERN announced~~\cite{Aubert:1983xm} what was an astonishing result at the time, namely that the $F_2$ structure function for deep inelastic scattering for Fe was very different in the valence region from that of D~\cite{Geesaman:1995yd}. Before the experiment, the fact that deep inelastic scattering is light-cone dominated meant that it was regarded as being dominated by perturbative QCD and hence insensitive to environment. Afterwards the thinking changed very quickly and it is now understood that each nucleus is a unique target and its structure function the result of non-perturbative QCD. The beauty of the EMC effect is that it does directly illustrate the change in the momentum distribution of the {\em quarks} as the nuclear environment varies. 
\begin{center}
	\begin{figure}[H]
		\centering
		\includegraphics[angle=0,width=0.9\textwidth]{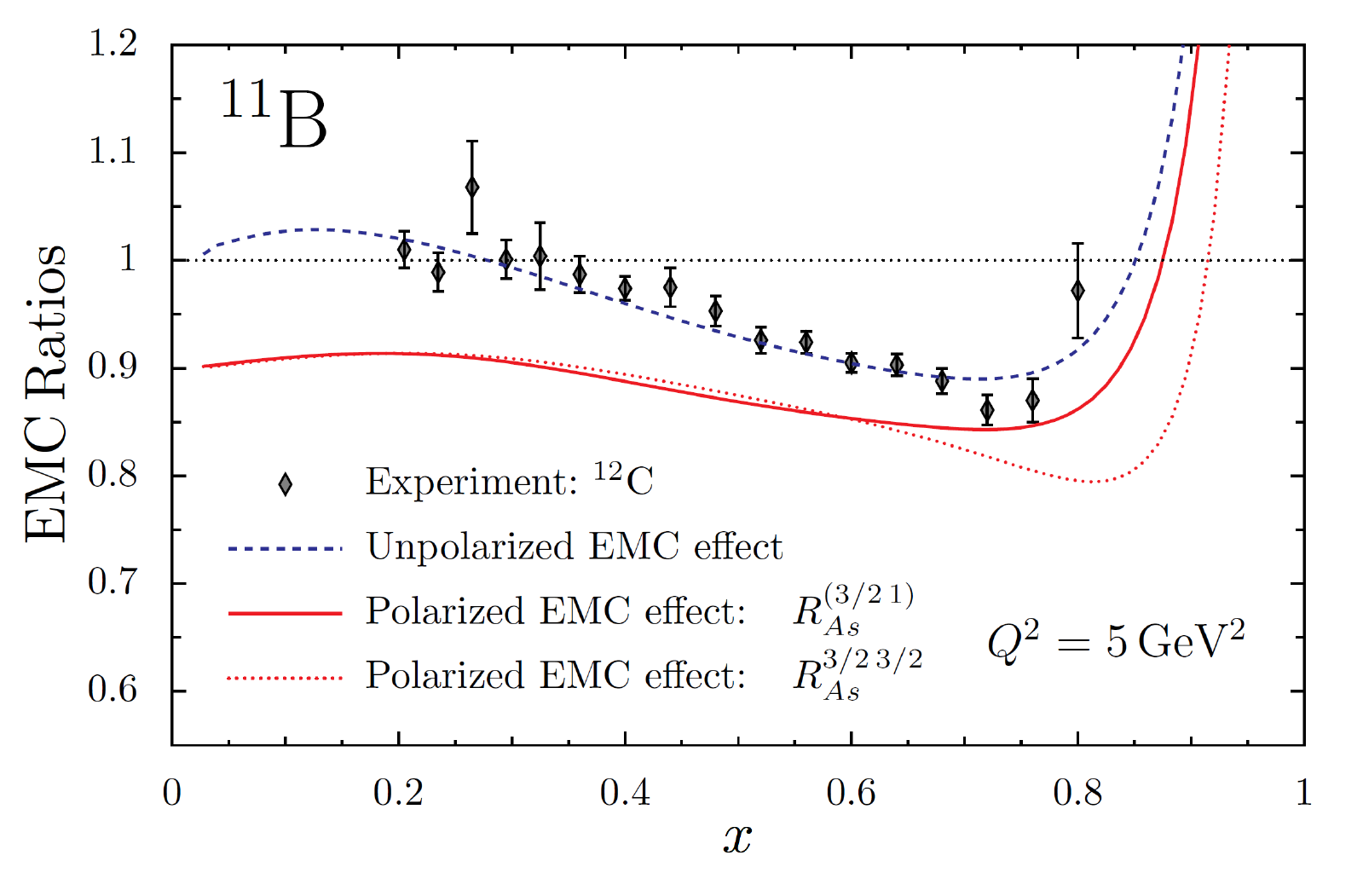}
		\caption{Comparison~\cite{Cloet:2006bq} of the data for the EMC effect in $^{12}$C, namely the ratio of the $F_2$ structure function for $^{12}$C to that of the deuteron,  with theoretical calculations of Clo\"et {\em et al.} (dotted line),  as a function of the light-cone momentum fraction, $x$, carried by the struck quark. (The calculation is labelled "Unpolarized'' to emphasise that it involves the structure function $F_2$.) The dotted and solid curves illustrate two polarized structure function observables for  $^{11}$B, divided by the effective polarization of the valence proton.  The deviation from unity is defined to be the polarized EMC effect. }
		\label{fig.B11}
	\end{figure}
\end{center}

Immediately following the EMC publication, a re-examination of SLAC data conformed the observed loss of momentum from valence quarks in other nuclei, with the size of the effect varying roughly as $A^{1/3}$ or the average nuclear density. Many experiments have since been conducted to probe the EMC effect 
further~\cite{Ashman:1988bf,Benvenuti:1987az,Gomez:1993ri}. Work at JLab clearly established that it was not the average density but the local nuclear density that mattered for the effect on valence 
quarks~\cite{Seely:2009gt}. 

In the present context we focus on the effect of the nuclear environment on the valence quark distribution. Another prominent feature of nuclear structure functions called shadowing is found at low $x$, in the region of the sea. There the interaction with one nucleon in the nucleus prevents the interaction with another so that the effective number of nucleons seen in the target is less than $A$. The physics associated with shadowing is fascinating but the origin appears to be well understood and quite separate from that associated with the EMC effect.

One of the first applications of the QMC model of nuclear structure was to nuclear structure functions~\cite{Saito:1992rm,Saito:1993yw} and indeed it was shown that the effect of the scalar and vector mean fields was consistent with the data. Such calculations have subsequently become much more sophisticated. In particular, in the work based upon the model of Nambu and Jona-Lasinio (NJL), which is covariant and respects chiral symmetry, where once again the relativistic mean scalar field modifies the internal structure of the bound nucleon. The fact that the model is covariant makes the calculation of parton distribution functions much simpler than in the bag-model. Figure~\ref{fig.B11} shows the excellent agreement between the data for the EMC effect on $^{12}$C, as an example,  and the 
calculations~\cite{Cloet:2006bq,Cloet:2005rt} based on the self-consistent change in the structure of the bound nucleon in this model.

\subsubsection{Polarized EMC effect}
In addition to describing existing data on the EMC effect, it is vital that any serious theoretical approach should generate further predictions which can be tested. This is especially important if there is more than one approach and in the case of the EMC effect that is the case. Data on high momentum nucleon knock-out in inelastic electron scattering has established the importance of the tensor force between neutron-proton pairs in generating high-momentum nucleons~\cite{Duer:2018sby}. The number of high-momentum pairs in a nucleus grows with the local density, like the EMC effect, and this has led to suggestions that they may be linked. If one nucleon from a highly correlated pair is emitted with high momentum, its partner will be far off-mass-shell. It has been argued that this may be the origin of the EMC effect~\cite{Schmookler:2019nvf}.

In an attempt to provide new predictions which can be tested, Clo\"et {\it et al.} 
calculated the nuclear modification of the spin-dependent structure function of a polarized nucleus, labelled the  ``polarized EMC  effect''~\cite{Cloet:2006bq,Cloet:2005rt} . 
This involves more difficult experiments because, in the ideal case, the nuclear polarization is carried by a single nucleon and the effect is therefore of order $1/A$. 
The predicted effect for $^{11}$B is illustrated in 
Fig.~\ref{fig.B11}, where we see that it is considerably larger than the unpolarized effect. A more recent calculation in the QMC model using the MIT bag to describe nucleon structure found a less dramatic effect with the unpolarized and the spin-dependent EMC effects roughly equal in size~\cite{Tronchin:2018mvu}. 

It is important to realize that the polarized effect is defined as the nuclear spin structure function divided by the effective polarization of the active proton multiplied by the free-proton spin structure function. Thus, only if the bound proton has a modified spin structure function will the ratio differ from unity. Of course, this does require that one has sufficiently good nuclear structure calculations that the effective polarization is known. Fortunately, for nuclei like $^{7}$Li and $^{11}$B, where the first experiments are likely to be carried out, this issue is under excellent control.

Only recently has it been realized just how important the polarized EMC effect, as defined above, is as a tool to distinguish between the QMC explanation and that based on short-range correlations. The nature of the tensor force is such that a polarized proton experiencing hard scattering with a neutron through the tensor force will essentially lose its polarization. Thus if the EMC effect arises only from highly correlated nucleons there will be no polarized EMC 
effect~\cite{Thomas:2018kcx}. This makes experiments such as that proposed at 
JLab~\cite{JLab}, which will establish whether or not there is {\em any} spin EMC effect, extremely important.

\subsubsection{Isovector EMC effect}
Yet another proposal to deepen our understanding of the EMC effect is the proposal, by 
Clo\"et {\em et al.}~\cite{Cloet:2009qs}, that in nuclei with more neutrons than protons the isovector nuclear force exerts repulsion on all down-quarks and attraction on all up-quarks. This has the effect of shifting light-cone momentum from {\em all} up-quarks to all down-quarks, regardless of whether the quarks are in a proton- or neutron-like cluster. This leads to a significantly larger EMC effect on up-quarks in a nucleus compared with that on the down-quarks. Although such an effect is completely consistent with isospin symmetry, it acts exactly like an increase in the level of charge-symmetry violation~\cite{Londergan:2009kj} for valence quarks. This is critical for an analysis of the NuTeV experiment, which reported a three standard deviation discrepancy for $\sin^2 \theta_W$ compared with the Standard Model~\cite{Zeller:2001hh}. The analysis relied on corrections to the Paschos-Wolfenstein result~\cite{Paschos:1972kj} for an isospin-zero target being small for iron. However, both charge-symmetry violation and the isovector EMC effect are non-negligible and act in the same direction, reducing the quoted anomaly to less than one standard deviation~\cite{Bentz:2009yy}.

There are very promising possibilities for testing the predictions of any theory proposed to explain the EMC effect in the isovector domain using parity-violating deep inelastic scattering~\cite{Cloet:2012td}. A future electron-ion collider may allow a direct measurement of the EMC effect on different quark flavors but for the present JLab offers the best possibilities for testing the predictions for the isovector EMC effect.  

\section{Conclusion}
\label{sec.conclusion}
The strong force that binds atomic nuclei is governed by the rules of Quantum Chromodynamics. There are strong arguments that suggest that the internal quark structure of a nucleon will adjust self-consistently to the local mean scalar field in a nuclear medium and that this may play a profound role in nuclear structure. There are promising attempts to calculate the properties of nuclei within lattice QCD but realistic nuclear-structure calculations based on lattice methods are many years away. For the present we are therefore led to work with quark-based models of hadron structure as the basis for building models of nuclear structure that account for this new physical insight.

With the derivation of a non-relativistic energy-density-functional based upon the quark-meson coupling (QMC) model, it is now possible to make competitive mean field theory calculations of nuclear properties across the periodic table. This has led to predictions for binding energies that are comparable with state-of-the-art Skyrme forces but with only half of the number of parameters. The predictions for superheavy nuclei, which were not part of the fit, were especially interesting, with the level of agreement with empirical binding energies at the 0.1\% level. 

The change in our physical picture of the atomic nucleus implied by this approach is fundamental, consituting a new paradigm for nuclear theory. Given its importance, it is crucial to find ways to test the prediction that the internal structure of a bound nucleon differs from that in free space. This is non-trivial, as by definition  a bound nucleon cannot be separated from the medium that binds it. Rather, the consequences of the underlying theory must be calculated as well as possible for phenomena that might a priori be suspected of being sensitive to such changes. We have described a number of examples of this kind, including precise measurements of the Coulomb sum rule and variations on the EMC effect.

\section*{Acknowledgments }
I would like to thank S. D. Bass, W. Detmold, P.Shanahan and C. Simenel for helpful comments concerning this manuscript. It is a pleasure to acknowledge the many collaborators who have contributed to my understanding of the issues presented here, including P.~A.~M.~Guichon, J.~Stone, K.~Saito, K.~Tsushima, G.~Krein, H.~Matevosyan, W. Bentz, I. Clo\"et, C. Simenel, S.~Antic, K.~Martinez and T.~Motta.
This work was supported by the University of Adelaide and by the Australian Research Council through Discovery Projects DP151103101 and DP180100497.



\begin{thebibliography}{99}
\bibitem{foot}
This theorem has the consequence that any correction associated with the symmetry violating quark mass difference must be proportional to the square of that difference.
\end{thebibliography}
\end{document}